\documentstyle[12pt,boxedminipage,epsfig,wrapfig,floatflt,shadow]{article}

\begin{document}

\title{
Helping science and engineering majors to reflect upon teaching and learning issues
}
\author{Chandralekha Singh, Laura Moin and Christian Schunn \\ Department of Physics and Astronomy,\\ Department of Psychology and 
Learning Research and Development Center
\\ University of Pittsburgh, Pittsburgh, PA, 15260}
\date{ }

\maketitle

\begin{abstract}
Recruiting and retaining highly qualified physics and physical science teachers is critical for maintaining
America's global competitiveness.
Unfortunately, only one third of the high school teachers in physics have a degree in physics and an even smaller 
number of physical science teachers in middle school have a good grasp of the scientific content they teach. 
Moreover, teachers often lack adequate pedagogical content knowledge to teach science effectively.
Here, we discuss the development, implementation, and assessment of a course for science and engineering undergraduates designed to
increase awareness and help them develop an interest and a deeper appreciation of the intellectual demands of physics teaching.
The course focused on increasing student enthusiasm and confidence in teaching by providing well supported teaching opportunities
and exposure to physics education research. The course assessment methods include 1) pre/post-test measures of
attitude and expectations about science teaching, 2) self and peer evaluation of student teaching, 3) content-based
pre/post-tests given to students who received instruction from the student teachers, and 4) audio-taped focus group
discussions in the absence of the instructor and TA to evaluate student perspective on different aspects of the course and its impact.
\end{abstract}

\section{Background}

In the recent report ``Rising above the gathering storm: energizing and employing America for a brighter economic
future", a panel of experts convened by the National Academies calls for immediate effort to
strengthen our scientific competitiveness.~\cite{rags} Indeed, educating students who are well-versed in science is critical
for preserving our economic competitiveness and leadership. 
Physical science lays the foundation for later high school science courses and an understanding of
physics helps students make sense of topics in other science fields.  Therefore, many scientists have proposed a K-12 science curriculum
with ``Physics First".~\cite{physfirst1,physfirst2,physfirst3,physfirst4}. If the ``Physics First" idea is increasingly
adopted in school districts nationwide, the need to recruit and retain well-qualified physics and physical science teachers will increase further.

Recent data from American Institute of Physics (AIP) Research Center~\cite{AIP} shows that there are approximately 23,000 high school physics 
teachers nationwide. Approximately 1200 new teachers teach physics each year out of which approximately 400 have a major or minor in physics.~\cite{AIP} 
The AIP Statistical Research Center 2000-2001 High School Physics Survey shows that $32\%$ of high school physics teachers are ``Specialists" in that
they have a physics degree and also have physics teaching experience, $40\%$ are ``Career" physics teachers in that they do not have a physics degree
but have extensive experience in teaching physics, and $28\%$ are ``Occasional" physics teachers in that they neither have a degree nor experience in
teaching physics.~\cite{AIP} What is even more troubling is that the teachers often lack adequate pedagogical
content knowledge to teach science.~\cite{peda1,peda2} 
It is vital to enhance effort to recruit highly qualified physics and physical science teachers and to carry out appropriate professional development 
and mentoring activities for in-service teachers in high schools and middle schools to ensure that the students they teach develop an appreciation and 
deep understanding of science and scientific method and are well-prepared for a high tech workplace.

Several programs have been highly successful in providing professional development activities for in-service physics teachers.
Since scientific inquiry is a sense-making endeavor, these approaches typically employ a research-based pedagogy in which students learn 
both science and scientific method simultaneously and are constantly engaged in the learning process. 
These successful approaches attempt to bridge the gap between the abstract nature of the laws of physics and the concrete physical situations
in which they are applicable. 
Hands-on and minds-on investigations are combined with appropriate use of technology and mathematical modeling to enhance student learning.
Students work with their peers and the instructor acts as
their guide to ensure that students build on their prior knowledge and get an opportunity to construct a robust knowledge structure.

For example, the Physics Teaching Resource Agents (PTRA) program~\cite{PTRA} initiated by the American Association of Physics Teachers (AAPT) in 1985
with support from the National Science Foundation and the American Physical Society (APS) is a leading in-service physics professional development program.
A professional development approach that has been used nationwide to train approximately 2500 physics and physical science teachers is
based upon the Modeling Instruction.~\cite{modeling} Modeling Instruction is a research-based approach for teaching science that was designated
one of the seven best K-12 educational technology programs out of 134 programs in 2000 by the US Department of Education. The Modeling
Instruction in Physics was designated in 2001 by the US Department of Education as one of two exemplary programs in K-12 Science Education
out of 27 programs evaluated. 
Another research-based approach that has been effective in preparing both the in-service and the pre-service teachers is based upon the Physics by Inquiry
curriculum developed by the University of Washington Physics Education group.~\cite{pbi}
The Activities Based Physics group has conducted very successful joint
professional development workshops for K-12 and college faculty members for more than a decade on a variety of pedagogical approaches
related to physics teaching.~\cite{priscilla}

Numerous remedies have been attempted to alleviate the shortage of well-trained physics and physical science teachers. Remedies 
range from national to local policies and programs and include such approaches as emergency 
certification and out-of-field assignment to fill vacancies; alternative certification programs to 
hasten licensing requirements and job placement; tapping nontraditional candidate pools such as 
paraprofessionals, retired military, or career changers; providing scholarships, signing bonuses, 
or student loan forgiveness; and establishing partnerships between school districts and teacher 
preparation institutions to meet staffing needs cooperatively~\cite{nsta,aaee,gafney,shugart,forcier,villegas,usdoe,usdoea}.
Each remedy has certain costs and some degree of 
success. However, many remedies must resort to back pedaling to meet content knowledge 
qualifications, calling back to the educational fold those who have already left, or investing in 
populations who fail to complete licensing requirements. 

One of the most accessible potential recruits are science and engineering undergraduates who have not yet completed their degree.
According to a longitudinal research study conducted by Seymore and Hewitt~\cite{seymore},  $20\%$ of science, engineering
and mathematics undergraduates 
at one time consider careers in math or science teaching, although less than $8\%$ of them hold to the career interest. 
One strategy to get more undergraduates interested in majoring in physics and in careers in physics teaching is revamping of the 
introductory physics courses.~\cite{lillian}
These courses are taken by most undergraduates interested in majoring in science and engineering and can provide an opportunity
to recruit more physics majors and more undergraduates with an interest in teaching physics. If these courses are not taught effectively,
we are unlikely to produce a higher percentage of undergraduates with interest in majoring in physics and in a career in physics teaching.~\cite{lillian}

In order to get the undergraduates excited about K-12 teaching, colleges and Universities must take responsibility for providing
the undergraduate students appropriate opportunity, guidance and support. 
A solid partnership between science and science education departments is a positive move in this direction.
Physics departments in some Universities have taken a lead role in working
with their Schools of Education to provide such opportunities to their undergraduates. For example, the UTeach program at The University
of Texas at Austin has been successful
in forging a partnership with the School of Education to provide a degree in science and a teaching certification simultaneously.~\cite{uteach}
Some member institutions of the PhysTEC program, which is a joint program of APS and AAPT, have been successful in increasing the number of 
undergraduates who go into K-12 teaching after graduation.~\cite{phystec} 
One feature of the PhysTEC program that has been promising is the Teacher In Residence (TIR) program in which a well-trained teacher acts as a 
Liasion between the University and the partnering school district.
Some of the PhysTEC institutions have a Learning Assistants program~\cite{la} that provides
undergraduate students opportunities as teaching assistants in college courses to cultivate their interest in teaching. 
Recently, a partnership of a large number of institutions called ``PTEC" has been
formed which provides a forum for exchanging ideas about physics and physical science teacher preparation via a yearly conference and a website.~\cite{ptec}
Other novel approaches such as involving science undergraduates as discussion leaders in museums is also being piloted to increase their interest in
teaching and to recruit them as K-12 teachers.~\cite{steinberg}

\vspace*{-.08in}
\section{Introduction}
\vspace*{-.08in}

Here, we discuss the development, implementation and assessment of a course called ``Introduction to physics teaching" for 
science and engineering undergraduates so that they would consider K-12 teaching as a potential career choice.
The course was designed to increase awareness and develop a deeper appreciation
about the intellectual demands of physics teaching. The course attempted to increase student enthusiasm and confidence
in teaching by giving them opportunity to design instructional modules in pairs and teach in authentic college
recitation classes two times during the semester. 
We provided significant scaffolding support and guidance during the development of the modules but gradually
decreased the guidance to ensure that students develop confidence and self-reliance.
The course strived to improve students' knowledge of 
pedagogical issues, familiarize them with cognitive research and its implication
for teaching physics, and included extensive discussions of physics education research including topics
related to active engagement, effective curricula, student difficulties in learning
different physics content, epistemological and affective issues. Special attention was paid to helping
students see the relevance of these discussions to actual classroom teaching and learning.

\vspace*{-.08in}
\section{Course Details}
\vspace*{-.08in}

The course has been taught twice with a total of 12 students. A majority of students were science and
engineering undergraduates (sophomores-seniors) with two masters students from the school of education.
The cumulative grade point average for the students was between 2.5 to 3.5. At least a B grade average in 
introductory physics I and II was mandatory to enroll which was a requirement imposed by the department of
physics because each student pair was required to conduct two college recitations.
% classes.

An initial survey in the first class period to the students enrolled in the class suggests that a majority 
of students had some kind of teaching experience. The most common teaching experience was tutoring in high
school. The survey responses suggest that students felt confident in teaching the subject matter they tutored
earlier. When asked to rank-order the main reasons for having taught in the past, the students cited ``curiosity"
followed by ``a sense of being good at it", followed by ``a desire to work with children", and
``giving back to the community".

The class met for three hours per week for a semester and students obtained three credits for it.
Students were assigned readings of one or two journal articles about teaching and learning each week. 
They submitted answers to the questions assigned about the readings and discussed
the articles in class each week.

We used a field-tested ``Cognitive Apprenticeship Model"~\cite{cog} of teaching and learning which has three major
components: modeling, coaching, and fading. Modeling in this context refers to the instructor demonstrating
and exemplifying the criteria of good performance. Coaching refers to giving students opportunity to
practice the desired skills while providing guidance and support and fading refers to weaning the support
gradually so that students develop self-reliance. In the modeling phase, students worked through and
discussed modules from an exemplary curriculum, Physics by Inquiry~\cite{inquiry}, in pairs. There was extensive
discussion of the aspects of the modules that make them effective and the goals, objectives, and performance
targets that must have lead to the development of those modules. In the coaching and fading phases,
the student pairs developed, implemented and assessed two introductory physics tutorials and related pre-/post-tests
with scaffolding support from the instructor, teaching assistant (TA) and peers. 

Students were allowed to choose their partner and they stayed with the same partner for both tutorials.
All student pairs designed the two tutorials on the same broad topics: DC circuits and electromagnetic
induction. Although all student pairs employed the tutorial approach to teaching, there was flexibility in how to design the
tutorial.
For example, one group successfully employed cartoons in their tutorials.
Also, students were free to choose the focus of their 25 minute long tutorial (10+15 minutes
were spent on the pre-test and post-test respectively). Each student group determined the goals and performance
targets for their tutorial which was then discussed during the class. This class discussion was very useful in
helping students realize that they needed to sharpen their focus for a 25 minute tutorial instead of covering every
concept in DC circuits or electromagnetic induction. A majority of the preliminary development of the tutorials and the 
accompanying pre-/post-tests took place outside of the class and students iterated versions of the tutorials
with the instructor and TA. Then, each pair tested their pre-/post-tests and tutorials on fellow classmates and
used the discussion and feedback to modify their tutorial. The peers were very conscientious about providing
comments on both the strengths and weaknesses of the tutorials.

\section{Course Evaluation}

\subsection{Evaluating Tutor Effectiveness}

The content-based pre-/post-tests accompanying the tutorials were given to the introductory physics students during the
recitation. The typical pre-/post-test scores were $40\%$ and $90\%$ respectively with a Hake normalized gain of 
$0.8$~\cite{hake}. We note that the pre-test refers to the test given after traditional classroom instruction but before the
tutorials.

\vspace*{-.2in}
\subsection{Evaluating Impact on Tutors}
\vspace*{-.08in}

We developed a teaching evaluation protocol based upon an existing protocol (RTOP)~\cite{rtop}
which includes 15 questions on a Likert scale designed to evaluate different
aspects of teaching. The 15 questions in the protocol were further divided into two parts: the first 7 questions were
related to content/lesson plans/class design and the other 8 questions dealt with the class activities during instruction.
The following are some items:
\begin{itemize}
\item Class content was designed to elicit students' prior knowledge and preconceptions and build new concepts from there.
\item The lesson was designed to engage students as members of a learning community: engaged in talk that builds on each other's
ideas, that is based on evidence and responds to logical thinking.
\item Instructional strategy included useful representational tools (for example, symbols, charts, tables, and diagrams).
\item The activity actively engaged and motivated students rather than having them be passive receivers.
\end{itemize}
Each student was required to observe and critique the instruction of at least one other
pair in each of the two rounds in addition to evaluating their own performance. All of the classroom teaching by the
students were videotaped. After each round, we discussed the teaching evaluations of each group in class to stress
the aspects of teaching that were good and those in need of improvement. We found that the student evaluation of other
pairs were quite reliable and consistent with the instructor and TA evaluation. Students did a good job evaluating
the positive and negative aspects of other group's instruction. However, self-evaluations were not reliable and students
always rated themselves highly. Students were told that their grades will depend only on the evaluation
conducted by the instructor and the TA and not on the self and peer evaluations and that the self and peer evaluations
were to help them learn to critique various aspects of instruction. The fact that students rated themselves
higher than others may be because they were worried that the evaluation may factor into their course grade.

There was a clear difference between different student pairs in terms of how effectively they helped the introductory
physics students work on the tutorials in groups. There was a strong correlation between the extent to which group work
was motivated and emphasized at the beginning of the recitation and its benefits explained and whether introductory students
worked effectively in groups. These issues were discussed with the student pairs, and each student pair
obtained a copy of all of their evaluations. They were asked to pay attention to the instructor/TA/peer critiques of their
performance. However, the second performance of each pair was not significantly different from the first. For example,
pairs good at employing group work effectively the first time did it well the second time and those who
had difficulty the first time had similar difficulties the second time. More detailed guidance is needed for 
improving students' classroom delivery methods.

We also conducted an anonymous survey in the absence of the course instructor at the end of the course. One of the questions
on the survey asked students to rate how the course affected their interest in becoming a teacher. $56\%$ reported a significant
positive impact, $34\%$ a positive impact and $10\%$ no impact. Students noted that they learned about the intellectual rigor of
instructional design from moderate to great extent. On a scale of 1 to 5, students were asked to rate different elements that
contributed to learning. They provided the following responses:
\begin{itemize}
\item Preparing tutorials and presentations: 4.8/5
\item Instructor's feedback on these: 4.5/5
\item Class discussions: 4.3/5
\item Rehearsals for their presentation: 4.0/5
\item Instructor's presentations: 4.0/5
\item Readings: 3.9/5
\end{itemize}

We also conducted an audio-taped focus group discussion to obtain useful feedback to evaluate and improve next offering of the
course. The focus group was conducted on the last day of class in the absence of the instructor and the TA. The facilitator
asked students pre-planned questions for one hour. The questions and some typical responses are presented below:

\underline{Question 1:} What is the take home message of this course?
\begin{itemize}
\item S1: Teaching is more than teacher's perception. How much of a two way relation is necessary to teach students.
\item S2: Helped me understand that teachers have to learn from students.
\item S3: Instruction is more students. There are methods available to make instruction more suited to students. There is
a mountain of cognitive research that is being developed as a resource for me as a future teacher...that was my biggest
fear when we started talking about bringing instruction to student's level.
\item S4: Increased enthusiasm. You have to take into account student's level.
\item S5: Increased appreciation of teaching. Opened my eyes to the difficulty and different techniques for teaching students 
with different prior knowledge.
\item S6: Figuring out different ways of making students active and structuring the lessons so that there is a lot of activity
by students to learn on a regular basis.
\end{itemize}
\underline{Question 2:} Do you take a different perspective during your own classes after you learned something about how to teach?
\begin{itemize}
\item S1: I think now that teachers who don't teach well could be trained but before the course I just took it for granted that
there are good and there are bad teachers and that's all.
\item S2: My college instructors ignore the work being done in how people learn.
\item S3: Slightly, because I know how difficult it is. I give more respect to good teachers.
\item S4: It gives you an idea about how a teacher cares about the students.
\end{itemize}
\underline{Question 3:} What did you learn from your K-12 teaching? How do you compare that to teaching 
at the college level?\\
A common response was that the students had not thought about teaching issues in high school or till they took this course.
\begin{itemize}
\item S1: When I was a student I just took teaching for granted and did what they told me to.
\item S2: I never thought about teaching when I was in high school.
\item S3: At school most were educators in college not.
\end{itemize}
\underline{Question 4:} How did this course affect your interest in teaching? 
What about your plans for pursuing teaching?\\
All students except two said they will teach. A majority explicitly said they plan to teach in high school. 
\begin{itemize}
\item S1: Reinforces my interest. Made me realize that I don't want to teach college because of the structure of college-lots of
material, little support, under-appreciated...I want to have more time to engage students in the method learned in this course. 
\item S2: It helped me decide I want to go on to teaching right after college.
\item S3: I want to be a teacher. This course affected me positively.
\item S4: K-12. Good physics teacher in high school to give good base at young age...early
\end{itemize}
\underline{Question 5:} How could this course be improved to enthuse more people to teaching?\\
One common discouraging response was that students felt they did not really get an opportunity
to teach where the word ''teaching" referred to frontal teaching. Despite the fact that the course
attempted to bridge the gap between teaching and learning, students felt that moving around the
classroom helping students while they worked on the tutorials was not teaching.
Common suggestions included a follow-up class with the following features:
\begin{itemize}
\item Observing, critiquing $\&$ delivering frontal teaching
\item Observing and critiquing K-12 teaching
\item Amount of reading per week can be reduced although students appreciated the readings
\end{itemize}

\vspace*{-.13in}
\section{Summary}
\vspace*{-.04in}

To prepare future scientists and engineers for the demands of a high tech workplace,
preparation of highly qualified K-12 science teachers is critica~\cite{1,2,3,4,5}. The physics departments in colleges and Universities must 
take responsibility to accomplish this important task.
We have developed, implemented and assessed a course for science undergraduates to increase their interest and awareness
about teaching issues. In addition to extensive discussions about issues related to teaching and learning, student
pairs designed and implemented two tutorials in college recitation classes. Assessment methods include pre-/post-tests
of expectation and attitude about teaching, content-based pre-/post-tests before and after tutorials designed by students,
critiquing peers and self-evaluation of teaching and focus group discussions.

\vspace*{-.1in}

\bibliographystyle{aipproc}  

\begin{thebibliography}{a}

\bibitem{rags} Rising above the gathering storm: Energizing and employing America for a bright economic future,
The National Academies Press, 2005.\\
Available online at http://www.nap.edu/books/0309100399/html
\bibitem{physfirst1} L. M. Lederman, ``Physics First?" The Phys. Teacher {\bf 43}, 6, (2005).
\bibitem{physfirst2} A. Hobson, ``Considering Physics First" The Phys.  Teacher {\bf 43}, 485 (2005).
\bibitem{physfirst3} O. Dreon Jr., ``A Study of Physics First Curricula in Pennsylvania" The Phys. Teacher {\bf 44}, 521, (2006).
\bibitem{physfirst4} B. Bessin, ``Why Physics First?" The Phys. Teacher, {\bf 45}, 134, (2007).
\bibitem{AIP} http://www.aip.org/statistics/
\bibitem{peda1} L. S. Shulman, Those Who Understand: Knowledge Growth in Teaching. Educational Researcher, 15(2), 4-14, (1986).
\bibitem{peda2} L. S. Shulman, Knowledge and Teaching: Foundations of the New Reform. Harvard Educational Review, 57(1), 1-22, (1987).
\bibitem{PTRA} http://www.aapt.org/Programs/projects/PTRA/
\bibitem{modeling} http://modeling.asu.edu/
\bibitem{nsta} National Science Teachers Association, NSTA releases nationwide survey of science 
teacher credentials, assignments, and job satisfaction. Arlington, VA, 2000: N.S.T.A. Retrieved from 
http://www.nsta.org/survey3/
\bibitem{pbi} Physics by Inquiry, L.C. McDermott and the Physics Education Group at the University of Washington, Volumes I and II,
Wiley, NY, 1996. 
\bibitem{priscilla} For example, see 
%http://physics.dickinson.edu/~wp_web/wp_resources/wp_workshops.html
\bibitem{aaee} American Association for Employment in Education, Inc., Educator Supply and Demand 
in the United States.  Columbus, OH: AAEE, 2000.  Retrieved from 
$www.rnt.org/channels/clearinghouse/becometeacher/121-teachershort.htm$
\bibitem{gafney} L. Gafney and M. Weiner, Finding future teachers from among undergraduate science and 
mathematics majors. Phi Delta Kappa, 76, 637-641, 1995.
\bibitem{shugart} S. Shugart and P. Houshell, Subject matter competence and the recruitment and retention 
of secondary science teachers. J. Research Science Teach., 32, 63-70, 1995.
\bibitem{forcier} B. C. Clewell, and L.B. Forcier, Increasing the number of math and science teachers:  A 
review of teacher recruitment programs. Teaching and Change,  8(4), 331-361, 2001.
\bibitem{villegas} B. C. Clewell, and A.M. Villegas, Absence unexcused: Ending teacher shortages in high-need 
areas. Washington, DC: Urban Institute. Retrieved from $www.urgan.org/url.cfm?ID=310379$, 2001.
\bibitem{usdoe} U.S. Department of Education, America's teachers:  Profile of a profession, 1993-94. 
NCES 97-460, 1997. Washington, DC: National Center for Education Statistics. Retrieved from 
$http://nces.ed.gov/pubsearch/pubsinfo.asp?pubid=97460$
\bibitem{usdoea} U.S. Department of Education, Teacher recruitment programs: 
Planning and Evaluation Service.  Office of the Under Secretary.  Washington, DC: The Urban Institute, 2000. 
\bibitem{lillian} L.C. McDermott, Guest Editorial, Preparing K-12 teachers in physics:  Insights from history, experience, and research, 
Am. J. Phys. 74 (9) 758-762 (2006).
\bibitem{seymore} E. Seymore and N. Hewitt, Talking about leaving: Why undergraduates leave the 
sciences.  Boulder, CO: Westview Press, 1997.
\bibitem{uteach} For example see http://uteach.utexas.edu/
\bibitem{phystec} For example see http://www.phystec.org/
\bibitem{la} V. Otero, N. Finkelstein, R. McCray and S. Pollock, Who is responsible for preparing science teachers?, Science,
{\bf 313}, 445-446, (2006).
\bibitem{ptec} For example, see http://www.compadre.org/ptec/ 
\bibitem{steinberg} For example, see information about CLUSTER (Collaboration for leadership in urban science teaching) at
http://www.sci.ccny.cuny.edu/~rstein/Cluster/cluster.html
\bibitem{cog} A. Collins, J. S. Brown, and S. E. Newman, {\it Cognitive Apprenticeship:
Teaching the crafts of reading, writing and mathematics}, 
in L. B. Resnick (Ed.), {\it Knowing, learning, and instruction:  Essays in honor of Robert Glaser}, 
Hillsdale, NJ:  Lawrence Erlbaum., 453-494, 1989.
\bibitem{inquiry} L. McDermott and PER group at university of Washington, Physics by Inquiry Vol I and II, Wiley 1996.
\bibitem{hake} R. R. Hake, 
{\it  Interactive-engagement versus traditional methods},
Am. J. Phys. {\bf 66}, 64 (1998).
\bibitem{rtop} http://physicsed.buffalostate.edu/AZTEC/RTOP/RTOP-full/index.htm
\bibitem{1} S. Shugart and P. Houshell, (2001). Subject matter 
competence and the recruitment and retention of 
secondary science teachers. J. Research Science 
Teach., 32, 63-70.
\bibitem{2} L. S. Shulman, (1986). Those who understand: 
Knowledge growth in teaching. Educational 
Researcher, 15(2), 4-14.
\bibitem{3} L. S. Shulman, (1987). Knowledge and teaching: 
Foundations of the new reform. Harvard 
Educational Review, 57(1), 1-22.
\bibitem{4} C. Singh and C. D. Schunn (2009). Connecting three 
pivotal concepts in K-12 science state standards 
and maps of conceptual growth to research in 
physics education, J. Phys. Tchr. Educ. Online, 
{\bf 5}(2), 16-42.
\bibitem{5} C. J. Wenning (2009), The new Aristotelianism?, 
Editorial, J. Phys. Tchr. Educ. Online, 5(2), 1-2.
\end{thebibliography}
{}
\end{document}